\documentclass[aps,twocolumn,nofootinbib,showpacs,floatfix]{revtex4}
\usepackage{graphics} 
\usepackage{amssymb,amsmath}
\usepackage{amsmath}
\usepackage{psfrag}
\usepackage{graphicx}

\def\be{\begin{equation}} 
\def\ee{\end{equation}} 
\def\bea{\begin{eqnarray}} 
\def\eea{\end{eqnarray}}  
\def\bean{\begin{eqnarray*}} 
\def\eean{\end{eqnarray*}}

\def\bse{\begin{subequations}}
\def\ese{\end{subequations}}

\def\lsim{\raise 0.4ex\hbox{$<$}\kern -0.8em\lower 0.62ex\hbox{$\sim$}} 
\def\gsim{\raise 0.4ex\hbox{$>$}\kern -0.7em\lower 0.62ex\hbox{$\sim$}}

\def\f0N{f_0^{(N)}}
\def\bec{\begin{center}}
\def\eec{\end{center}}

\begin{document} 
\title{Quasi-stationary states and the range of pair interactions}

\author{A. Gabrielli$^{1,2}$, M. Joyce$^{3,4}$ and B. Marcos$^{5}$} 

\affiliation{$^1$SMC, CNR-INFM, Physics Department, University 
``Sapienza'' of Rome, Piazzale Aldo Moro 2, 00185-Rome, Italy}
\affiliation{$^2$Istituto dei Sistemi Complessi - CNR, Via dei Taurini 19, 
00185-Rome, Italy}
\affiliation{$^3$Laboratoire de Physique Nucl\'eaire et Hautes \'Energies,\\
Universit\'e Pierre et Marie Curie - Paris 6,
CNRS IN2P3 UMR 7585, 4 Place Jussieu, 75752 Paris Cedex 05, France}
\affiliation{$^4$Laboratoire de Physique Th\'eorique de la Mati\`ere Condens\'ee,\\
Universit\'e Pierre et Marie Curie - Paris 6,
CNRS UMR 7600, 4 Place Jussieu, 75752 Paris Cedex 05, France}
\affiliation{$^5$Laboratoire J.-A. Dieudonn\'e, UMR 6621, 
Universit\'e de Nice --- Sophia Antipolis,
Parc Valrose 06108 Nice Cedex 02, France} 

\begin{abstract}   
\begin{center}    
{\large\bf Abstract}
\end{center}    
``Quasi-stationary'' states are approximately time-independent 
out of equilibrium states which have been observed in a variety of systems
of particles interacting by long-range interactions. We investigate here
the conditions of their occurrence for a generic pair interaction 
$V(r \rightarrow \infty) \sim 1/r^\gamma$  with $\gamma > 0$,
in $d>1$ dimensions. We generalize analytic calculations
known for gravity in $d=3$ to determine the scaling parametric
dependences of their relaxation rates due to two body collisions, and 
report extensive numerical simulations testing their validity. 
Our results lead to the conclusion that, for $\gamma < d-1$,
the existence of quasi-stationary states is ensured by the 
large distance behavior of the interaction alone, while for $\gamma > d-1$
it is conditioned on the short distance properties of the interaction,
requiring the presence of a sufficiently large soft core in the interaction potential.
\end{abstract}    
\maketitle   
\date{today}  

\twocolumngrid   

In recent years there has been renewed interest in the statistical
physics of long-range interactions (for a review, see e.g. \cite{campa_etal_LRreview_2009}), 
a subject which has been treated otherwise mostly in the astrophysical
literature for the specific case of gravity.
The defining property
of such interactions is the non-additivity of the potential
energy of a uniform system, which corresponds to the non-integrability
at large distances of the associated pair interaction, i.e., a
pair interaction $V (r \to \infty) \sim 1/r^\gamma$ with 
$\gamma < d$ in $d$ space dimensions. The equilibrium thermodynamic 
analysis of these systems is very different to the canonical one for 
short-ranged interactions (with $\gamma > d$), leading notably
to inhomogeneous equilibria as well as other unusual properties ---
e.g. non-equivalence of the statistical ensembles, negative specific 
heat  in the microcanonical
ensemble.
Studies of simple toy models have shown that, like for
gravity in $d=3$, these equilibria (when defined) are reached
only on time scales which are extremely long compared to those
characteristic of the mean field dynamics. On the latter time scales 
one observes typically the formation, through ``violent  relaxation",
of so-called ``quasi-stationary'' states (QSS), 
interpreted theoretically as stable stationary states of the 
Vlasov equation (which describes the kinetics in the mean field limit).
%
%
In this letter we consider
whether the occurrence 
of such QSS driven by mean-field dynamics can be considered
as a behavior arising generically when there are long-range 
interactions in play. 
%
Using 
both 
simple analytical 
results and 
numerical simulations, we argue
for the conclusion that it is only for $\gamma < d-1$, 
i.e. when the pair {\it force} is absolutely integrable at large
separations, that QSS can be expected to occur
independently of the short distance properties of the interaction.
For $\gamma > d-1$, on the other hand, their occurrence will
be conditioned strongly also on short distance properties, and 
thus cannot be considered to be a result simply of the long-range
nature of the interaction.  Our analysis shows the relevance
of a classification of the range of interactions according to the 
convergence properties of {\it forces} rather than  
potential energies  which has been formalized in 
\cite{dynamical_classification_2010}.

We proceed 
by first generalizing a calculation originally given by
 Chandrasekhar for Newtonian gravity to a
 system of $N$ particles interacting by a pair potential 
 $V(r)=\frac{g}{r^\gamma}$ (where $g$ is a 
 coupling constant).  This calculation, which numerical 
studies indicate is accurate both parametrically and 
quantitatively for gravity (see e.g.  \cite{binney, theis_1998, theis+spurzem_1999, diemandetal_2body, levin_etal_2008}),
will give us an estimate of  $\Gamma_{2}$, 
the relaxation rate due to two body collisions (i.e. the inverse
of the time-scale on which a typical particle's velocity is 
randomized by such interactions).  
Denoting by $\tau_{mf}$ the characteristic time for 
the formation of a QSS (i.e. of the mean-field dynamics),
the criterion for the existence of QSS we will then study 
is 
\be
\Gamma_{2} \, \tau_{mf} \to 0 \quad {\rm when} \quad N \to \infty\,, 
\label{criterion-QSS}
\ee
where the limit $N \to \infty$ 
corresponds to the mean-field or Vlasov limit \cite{campa_etal_LRreview_2009}.
Indeed if this condition is not satisfied, it implies that there is no mean-field regime 
in which QSS may form.



Following the treatment for the case of gravity  (see e.g. \cite{binney},  section 1.2.1) 
we consider a test particle of velocity $v$ crossing a system in a QSS, 
assumed spherical and of radius $R$ and approximated as homogeneous.
We estimate first the rate of
relaxation due to  {\it soft} two-body collisions by calculating $\Delta v^2$, {\it the 
mean square velocity change of a particle per crossing}
(i.e. in a time of order $\tau_{mf}$) due to such collisions.
It is straightforward to show that
\be
\frac{\Delta v^2}{v^2} \sim N \, \left(\frac{g}{mv^2 R^\gamma}\right)^2 \int_{b_{min}/R}^{b_{max}/R}
\frac{dx}{x^{2\gamma-d+2}}\,,
\label{delta-v-squared}
\ee
where $b_{min}$ is the minimal impact parameter at which the scattering 
is soft (i.e. the deflection angle
is small), defined by 
\be
\frac{|g|}{mv^2 b_{min}^\gamma}  \sim  1\,,
\label{weak-deflection-condition-b}
\ee
and $b_{max}$ is the maximal impact parameter for two body collisions.
In these formulae, and in what follows below, we use the symbol  $\sim$ to 
indicate that 
the numerical factors  in all expressions
have been dropped, leaving only the parametric dependences
which are relevant to our considerations here.  In the case of gravity in $d=3$ the 
choice of $b_{max}$ has been a source of debate, with numerical simulations
 indicating 
that $b_{max} \sim R$ accounts better for results than the  
more evident choice  $b_{max} \sim \ell$, the mean inter-particle 
separation (see e.g. \cite{theis_1998}). We consider in what
follows both possibilities, and will see that  our central results
are not in fact sensitive to which is correct. 
We have also implicitly assumed $d>1$ and  $\gamma >0$.

We now write
\be
\Gamma_{2}\, \tau_{mf} = \Gamma_{soft} \,\tau_{mf} +\Gamma_{hard} \,\tau_{mf}, 
\label{2scattering-decomposition}
\ee
where the first contribution is that considered above, 
and the second is the remaining one from hard scatterings, i.e., 
collisions with impact factors $ b< b_{min}$.  Taking now
that $\tau_{mf} \sim \frac{R}{v}$, it is straightforward to deduce
from Eq.~(\ref{delta-v-squared}) that, for sufficiently
large $N$,  
\be 
\Gamma_{soft} \,  \tau_{mf}  \sim  \left\{
 \begin{array}{cc}
     N^{-1}  \, \left(\frac{b_{max}}{R}\right)^{-2\gamma+d-1}  & \,{\rm if}\, \gamma  < (d-1)/2  \\ 
     N^{-1} \, \left(\frac{R}{b_{min}}\right)^{2\gamma-d+1}    &  \,{\rm if}\,  \gamma  > (d-1)/2  \\ 
  \end{array}\right. \,,
\label{rate-soft-core}
\ee 
 if $b_{min}/b_{max} \ll 1$ for large $N$. To infer these scalings we need only 
 (as in the corresponding derivation for the case of gravity \cite{binney}) use
 the fact that the QSS is, by definition, a virialized state, i.e., we take 
\be
\frac{g}{mv^2 R^\gamma} \sim \frac{1}{N} \frac{gN^2}{(mNv^2) R^\gamma} 
\sim \frac{1}{N} \frac{U}{K} \sim \frac{1}{N}
\label{scaling-virialization}
\ee
where $U$, the total potential energy of the QSS,  and $K$,  its total kinetic energy, 
have a fixed ratio because of virialization. 
This scaling with $N$ corresponds to that in the usual mean field 
or Vlasov limit, in which $U$ and $K$ 
both scale in the same way with $N$.

Using again the scaling  Eq.~(\ref{scaling-virialization}),  the definition
Eq.~(\ref{weak-deflection-condition-b}) gives
\be
b_{min}  \sim  R N^{- \frac{1}{\gamma}}\,.
\label{condition-soft-scattering-1}
\ee
Note firstly that this implies  $b_{min}/b_{max} \to 0$ as $N \to \infty$ 
for any $\gamma>0$  if $b_{max} \sim R$, and for any  
$0 < \gamma < d$ if $b_{max} \sim \ell \sim RN^{-1/d}$,
so that Eq.~(\ref{rate-soft-core}) is indeed valid in these cases. 
Using now again Eq.~(\ref{condition-soft-scattering-1}) in 
Eq.~(\ref{rate-soft-core}) we obtain the scaling 
\be 
\Gamma_{soft} \,  \tau_{mf}  \sim  \left\{
 \begin{array}{cc}
     N^{-(1+|\delta|)}   & \,{\rm if}\, \gamma  < (d-1)/2   \\ 
     N^{-(d-1-\gamma)/\gamma}     &  \,{\rm if}\,  \gamma  > (d-1)/2  \\ 
  \end{array}\right. \,,
\label{rate-epsilon=0}
\ee
where $\delta=0$ if $b_{max} \sim R$, and $\delta =(-2\gamma+d-1)/d$
if $b_{max} \sim R N^{-1/d} $.
It follows that,  for $\gamma > d-1$,  the contribution of soft two
body scatterings alone diverges at large $N$, so that the criterion
(\ref{criterion-QSS}) cannot be satisfied in
this case for
the ``candidate" QSS.  
For any $\gamma < d-1$, on the other hand, the contribution 
$\Gamma_{soft} \,  \tau_{mf}$  vanishes as $N \rightarrow \infty$.
It is simple to show, in this case, that   $\Gamma_{hard} \,  \tau_{mf}$  also 
goes  to zero when $N \rightarrow \infty$, and thus that the
condition (\ref{criterion-QSS}) for the existence of QSS 
may be satisfied. 
To do so it is sufficient to consider 
that this contribution can be {\it bounded below} by that from 
an ``exactly hard'' core with radius   $\epsilon=b_{min}$, i.e., 
$V(r)=\infty$ for $r < b_{min}$.  Estimating the 
collision rate on such a core as $\Gamma_{hc} \sim n \sigma v$
where $n$ is the mean density and 
$\sigma \sim \epsilon^{d-1}$ (the cross section), we obtain
\be
\Gamma_{hc} \, \tau_{mf}  \sim N \left( \frac{\epsilon}{R} \right)^{d-1}  \sim N^{-(d-1-\gamma)/\gamma} 
\label{rate-hard-core}
\ee
when we take $\epsilon=b_{min}$, with the latter scaling as in  
Eq.~(\ref{condition-soft-scattering-1}). It follows that 
$\Gamma_{hard} \,  \tau_{mf} \leq \Gamma_{hc} \,  \tau_{mf} 
\to 0$ as $N \rightarrow \infty$ for $\gamma < d-1$. Further it follows 
from the inferred scaling of  $\Gamma_{hc} \,  \tau_{mf}$ that, for 
$\gamma < d-1$, the total rate $\Gamma_{2}$ will
scale as calculated for $\Gamma_{soft}$ in 
Eq.~(\ref{rate-epsilon=0}). In other words, an exact calculation including
$\Gamma_{hard}$ should give, at most, a $\Gamma_{2}$  larger 
than $\Gamma_{soft}$ by a numerical factor.

\begin{table}
\label{table1}
\caption{Summary of two body collision rates (without core)}
\begin{tabular}{|c|c|}
\hline
$0 < \gamma < \frac{d-1}{2}$ & soft collisions at $\sim b_{max}$ dominate \\
&$\Gamma_{soft} \,\tau_{mf} \sim  N^{-(1+|\delta|)} \gg \Gamma_{hard} \,\tau_{mf}$ \\
\hline
$ \frac{d-1}{2} < \gamma <  d-1$ & collisions at $\sim b_{min}$ dominate \\
&$\Gamma_{soft} \,\tau_{mf} \sim  N^{-\frac{d-1-\gamma}{\gamma}} \geq  \Gamma_{hard} \,\tau_{mf}$ \\
\hline
$ \gamma > d-1$ & $\Gamma_{soft} \,\tau_{mf}$ and $\Gamma_{hard} \,\tau_{mf}$ divergent in $N$\\
\hline
\end{tabular}
\end{table}

A corollary of these results, which are summarised in Table~\ref{table1}, 
is that, for a QSS to exist in the case that $\gamma > d-1$, the pair potential 
must  include a {\it sufficiently 
large
soft} core. Indeed to remove the divergence  of 
$\Gamma_{soft} \,  \tau_{mf}$ in this case, we must introduce a 
smoothing of the potential at a scale $\epsilon$ which {\it vanishes more slowly 
than} $b_{min}$ in Eq.~(\ref{condition-soft-scattering-1}).
In this case  $\Gamma_{soft} \,  \tau_{mf}$ is given by the 
second expression in Eq.~(\ref{rate-soft-core}) but with 
$b_{min}$ replaced by $\epsilon$. Keeping $\epsilon/R$
constant, for example, gives 
$\Gamma_{soft} \,  \tau_{mf} \sim N^{-1} \to 0$ as $N \to \infty$
for any $\gamma$.
If the core is ``exactly soft'', i.e., 
$V(r)=g/\epsilon^\gamma$ for $r<\epsilon$
we have $\Gamma_{hard}=0$ and the satisfaction of the condition
Eq.~(\ref{criterion-QSS}) follows. If the core is 
hard, as envisaged above, it is clear that
the same is not true. Indeed it is simple to check
using Eq.~(\ref{rate-hard-core}) that  it is not possible
to choose $\epsilon$ in order to satisfy 
both $\Gamma_{hc} \, \tau_{mf} \to 0$ 
and $\Gamma_{soft} \, \tau_{mf}  \to 0$ simultaneously
as $N\to \infty$ for $\gamma > d-1$. 

These results lead then to the primary conjecture of this article:  
for pair potentials with $V (r \to \infty) \sim 1/r^\gamma$, QSS can
always exist if there is a sufficiently large soft core, but 
only for $\gamma <  d-1$ can they exist when such a core
is not present (i.e. when its size $\epsilon \rightarrow 0$). 
The validity of this conclusion rests evidently on the assumption that 
the dominant correction to the mean-field dynamics is, just as for gravity in $d=3$,
two-body collisionality. More specifically we also require the parametric 
dependences of the inferred relaxation rates, which
have been derived using various simplifying approximations (notably 
that of homogeneity both in configuration and velocity space). We now 
present results of numerical simulations  (in $d=3$) which test their validity.  
We focus here on the crucial result above: the parametric dependence 
of the two-body  scattering rate due to soft scatterings in Eq.~(\ref{rate-soft-core}), 
for the range $\gamma > (d-1)/2$.

We perform molecular dynamics simulations using a version of the publicly
available gravity code GADGET2\cite{gadget}. We have modified the force 
routine in the tree-PM version of the code to treat a generic power-law pair 
potential with a core.
As in the original code we use a soft repulsive
core, with compact support: for $r<\epsilon$ $V(r)$ 
decreases continuously to a 
minimum at $r \approx \epsilon/2$ and then increases back to a local 
maximum $V(r=0)=0$.
In what follows the values of $\epsilon$ quoted correspond to the separation
at which the force is still attractive but has dropped to approximately 
$30\%$ of its value in absence of smoothing.
We consider here the attractive case (i.e. $g <0$).
The simulations are checked using simple convergence tests on the
numerical parameters, and their accuracy is monitored using 
energy  conservation. For the time-steps used here it is typically of order 
$0.1\%$ 
over the whole run, orders of magnitude smaller than the typical variation of 
the kinetic or potential energy over the same time. 
As initial conditions we take the $N$ particles on the sites of 
a simple cubic lattice of side $L_0$ and ascribe random velocities 
uniformly distributed in an interval $[-\Delta,\Delta]$ in each direction (i.e.
``waterbag" type initial conditions in phase space). The parameter
$\Delta$ is chosen so that initial virial ratio is unity, i.e.,
$2K/|U|=\gamma$. We choose this initial condition because it
would be expected to be close to a QSS to which (collisionless) 
relaxation should occur ``gently".  The system is enclosed in a 
cubic box of side $L \approx 2L_0$ (and centered on the same
point as the initial cube of particles).  Energy conserving ``soft" reflecting 
boundary conditions are used in the  dynamics, i.e., at each time step particles 
which have moved outside the box have the appropriate components 
of their velocity inverted.  The results we report here required
runs lasting as long as two weeks on up to sixteen processors.

\begin{figure}
\begin{center}
{\includegraphics*[width=8.5cm]{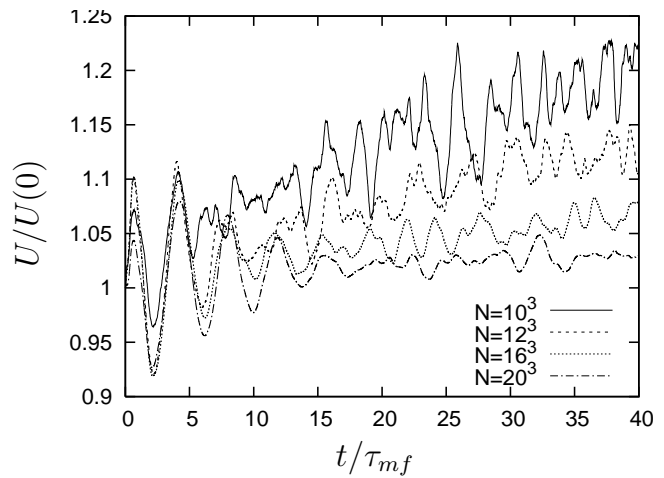}}
\caption{Temporal evolution of the total potential energy 
$U$ divided by its initial value $U(0)$, for $\gamma=1.25$ and a soft 
core $\epsilon/L=0.01$, for the different $N$ indicated.
}
\label{fig1}
\end{center}
\end{figure}
Shown in Fig.~\ref{fig1} is the evolution of the potential
energy $U$ as a function of time, for a pair potential with
$\gamma=5/4$ and $\epsilon/L=0.01$,  for the different values 
of  $N$ indicated. We have defined 
$\tau_{mf}= \sqrt{\frac{m L_0^{\gamma+2}}{gN}}$,
which, given that $L_0 \sim R$, is equivalent parametrically 
to the definition used above, assuming the scaling in
Eq.~(\ref{scaling-virialization}).
The macroscopic behavior monitored in this plot is clearly
very consistent with what has been anticipated, in line
with the typical behavior observed in self-gravitating
systems and other systems with  long-range
interactions studied in the literature:  there is 
a first phase of  ``violent'' (collisionless) relaxation towards an 
approximate  equilibrium, the QSS, which then evolves itself
in a second phase on a time-scale which clearly depends 
on $N$. The first phase, on the other hand, 
should be $N$-independent: as $N$ increases we see that
the different curves are increasingly well superimposed at 
early times. 

\begin{figure}
\begin{center}
{\includegraphics*[width=8.5cm]{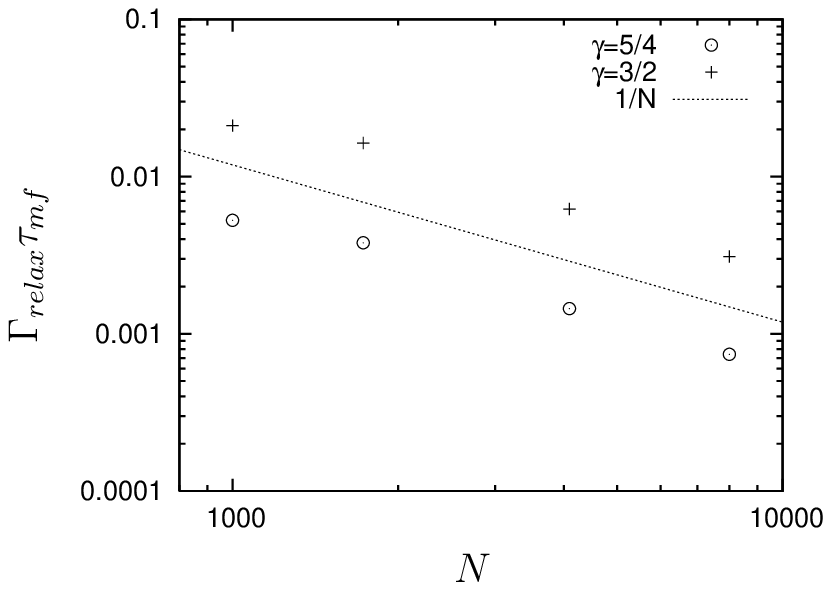}}\\
{\includegraphics*[width=8.5cm]{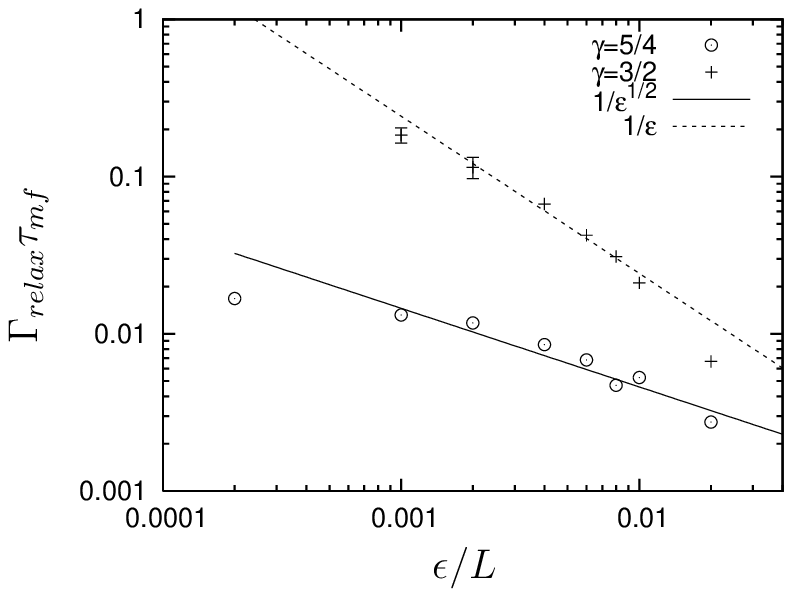}}
\caption{Estimated relaxation rate $\Gamma_{relax}$
as a function of $N$ at fixed  $\epsilon/L=0.01$ (upper panel), 
and as a function of $\epsilon$ at fixed  $N=10^3$ (lower panel).}
\label{fig2}
\end{center}
\end{figure}
Shown in Fig.~\ref{fig2} are, for the two 
cases $\gamma=5/4$ and $\gamma=3/2$, our
measurements of the relaxation rate $\Gamma_{relax}$, as a 
function of  $N$ (upper panel) at a chosen fixed $\epsilon$,
and as a function  $\epsilon$ (lower panel) at fixed chosen $N$.
The estimate of $\Gamma_{relax}$ is obtained 
simply from the slope of the potential energy  plotted as a 
function of time, in the  region in each case where this is well fit by 
a linear behavior, i.e.,  we take $\Gamma_{relax}=d(\ln U)/dt$  at 
$t \to 0$. Each point corresponds to one numerical simulation.
Note that for these determinations we consider thus only the
evolution away from, but still close to, the QSS.  Further 
results on the longer time evolution of these systems,
and in particular the compatibility of the fully relaxed
states with those predicted analytically for this case 
in  \cite{ispolatov_cohen}  (and related numerical studies in
\cite{ispolatov+karttunen_2004}) will be given elsewhere.

The upper panel of Fig.~\ref{fig2} includes a line showing 
the scaling proportional to $1/N$ predicted by 
Eq.~(\ref{rate-soft-core}) at fixed $b_{min}=\epsilon$.
The agreement is clearly very good.  Further it is simple
to verify that the results are {\it quantitatively} very coherent
with the prediction: taking $R \approx L_0/2 \approx L/4$, 
Eq.~(\ref{rate-soft-core}) fit the normalizations of
the plot with a prefactor of order unity in both cases.
While the degree of this concordance
--- despite the many approximations 
which inevitably limit the accuracy we can expect, and
the fact that we have dropped all numerical factors
in our derivation ---  is clearly just fortuitous, this quantitative
coherence of the results confirms their solidity.

The lower panel of Fig.~\ref{fig2} shows likewise
excellent agreement  with the predictions above. On it
are shown lines corresponding to the behavior of 
Eq.~(\ref{rate-soft-core}) at fixed $N$, when we
replace $b_{min}$ by $\epsilon$.  
As discussed above, this scaling is predicted to be
valid in the regime $b_{min} < \epsilon < b_{max}$.
Below $b_{min}$ we expect the rates 
to reach an asymptotic $\epsilon$-independent
value of order those estimated in Eq.~(\ref{rate-epsilon=0}).  
The behaviors in the plot are very coherent with these 
predictions, for values of $b_{min}$ which
are in good agreement with Eq.~(\ref{condition-soft-scattering-1}), 
taking again $R\approx L/4$. While we have not predicted the 
value of $b_{max}$, the downward deviation (corresponding
to a reduction in scattering rate) from the fit
at larger $\epsilon$ occurs at a value very consistent 
in each case with the measured mean inter-particle 
distance.

We note that this last plot explains why we consider 
only $\gamma$ up to $\gamma=3/2$, and indeed  
why we have not tried to verify more directly the
scalings in Eq.~(\ref{rate-epsilon=0}) using
simulations with $\epsilon \to 0$. The reason is
that, in order to measure the relaxation rates, we need 
to access the regime $\Gamma_{relax}\tau_{mf} \ll 1$,
i.e., we need to have a reasonable separation 
between the times scale of the collisionless dynamics
(and formation of QSS!) and the relaxation time scale.
At $N=10^3$ we see that $\gamma=3/2$ is already at 
this limit for the smallest $\epsilon$, and the error bars
on these points reflect the greater difficulty we have in 
making the measurement in these cases. The only remedy 
is to increase $N$, which, however, is prohibitively expensive
numerically, in particular at small $\epsilon$ where
the proper integration of the (few) hard collisions 
included requires significant decrease in the time
stepping. 

Finally a few remarks on the relation of these results
to some of the extensive recent literature on QSS
(see \cite{campa_etal_LRreview_2009} for references).
The determination of the $N$ dependence of QSS lifetimes 
has been much emphasized, both as a target for 
phenomenological studies of toy models, and for
theoretical studies of  the problem. 
Our results 
show that such lifetimes 
can be expected to depend, in general,  not just on $N$, 
but also on the parameters characterizing the short distance 
properties of the potential.
While for $\gamma < d-1$ a limit $\epsilon=0$ may be
defined [and gives the scaling of Eq.~(\ref{rate-epsilon=0})], 
for $\gamma > d-1$ this is not possible and the scaling of
the relaxation rate will depend necessarily on how $\epsilon$ 
scales with $N$. 
It would be interesting to extend our numerical simulations to 
explore the robustness of QSS notably to effects which may
come into play in more physically realistic settings: as 
shown by a recent study  \cite{gupta+mukamel_prl_2010}
of a toy model, the introduction of stochasticity in the dynamics
may also destroy QSS. 
We emphasize that our results here apply only to the particular
(albeit broad) class of models considered, and {\it a priori}
not, e.g.., to long-range spin models in which there is no
equivalent of two body collisions. It remains an interesting open
question to determine in a broader such context the conditions 
for the existence of QSS on the spatial dependence of the interaction.   


We acknowledge many useful discussions with F. Sicard. This work was partly 
supported by the ANR 09-JCJC-009401 INTERLOP project.


\end{document}